\documentclass[12pt]{iopart}
\begin{document}

\title[Spiked oscillators]{Spiked oscillators: exact solution}
\author{F J G\'omez and J Sesma}
\address{Departamento de F\'{\i}sica Te\'orica, Facultad de
Ciencias, 50009, Zaragoza, Spain}

\ead{javier@unizar.es}

\begin{abstract}
A procedure to obtain the eigenenergies and eigenfunctiones of a quantum spiked oscillator is presented. The originality of the method lies in an adequate use of asymptotic expansions of Wronskians of algebraic solutions of the Schr\"odinger equation. The procedure is applied to three familiar examples of spiked oscillators
\end{abstract}

\pacno{03.65.Ge, 02.30.Hq}




\section{Introduction}

Spiked harmonic oscillators, i. e., harmonic
oscillators to which a singular repulsion at the origin
\[
\frac{\lambda}{r^\alpha}, \qquad \lambda >0, \quad \alpha >0,
\]
has been added, have deserved a considerable interest since the
publication of the pioneering paper of Klauder \cite{klau},
three decades ago. A very extensive list of
articles dealing with the topic can be found in more recent papers by
Saad, Hall and Katatbeth \cite{saad} and by
Liverts and Mandelzweig \cite{live}.

Singular (at the origin) potentials were firstly considered in the
context of collision of particles \cite{fubi}, the main interest being
the adequate definition of the $S$ matrix. Here we are concerned with
the determination of the energy levels and the corresponding wave functions
of a particle bound in a confining potential behaving at infinity like
$r^n$, $n\geq 2$, and presenting a singularity of the type $r^{-m}$, $m>2$,
at the origin.

Besides of numerical integration of the Schr\"odinger equation by procedures
adapted to the singular potential \cite{kors,roy,buen} and iterative methods of the
Lanczos type \cite{sola},
different approximate methods have been implemented  to
solve spiked oscillator problems.
The usual WKB method was discarded from the very beginning in view of
the results of Detwiler and Klauder \cite{detw} which also showed that conventional
perturbative methods could not be used in the case of supersingular potentials.
Harrell, in a very lucid paper \cite{harr}, suggested a modified perturbation
theory to a finite order applicable in that case. A perturbative study of nonsingular
($\alpha < 5/2$) spiked harmonic oscillators in the two regimes of weak coupling
($\lambda\ll 1$) and strong coupling ($\lambda\gg 1$) has been carried out by
Aguilera-Navarro {\it et al} \cite{agui1}. They and other authors \cite{este}
have discussed the connection between the expansions obtained in the two regimes.
For the case of critical ($\alpha =5/2$) singularity, a large coupling perturbative
expansion has been obtained \cite{agui2}. One or more supersingular ($\alpha >5/2$)
terms added to the harmonic oscilator potential have been treated also perturbatively
by Guardiola and Ros \cite{guar} and by Hall and coworkers \cite{hall1,hall2}.
Nonperturbative procedures have also been used: let us mention different implementations
of variational methods \cite{saad,agui2,hall2,fern,hall3,agui3}, matricial \cite{znoj1} and
Hill-determinant \cite{znoj2} techniques, a conveniently modified WKB method \cite{tros},
smooth transformations of solvable potentials \cite{hall4}, the pseudoperturbative shifted-$l$
expansion technique \cite{must}, and the more recently proposed quasilinearization method \cite{live}.

All these procedures
provide with reasonably accurate values of the energies, at least
for the ground state, but are not able to give the wave function
in the vicinity of the origin. This fact, in addition to being
unsatisfactory from an aesthetical point of view, may lead to
incorrect expected values of certain operators.

The first attempt to obtain the correct wave functions of spiked
harmonic oscillators was done, to our knowledge, by Znojil
\cite{znoj3,znoj4}, who looked for solutions of the Schr\"odinger
equation in the form of Laurent series multiplied by a non integer
power of the variable. Substitution of such series in the
differential equation leads to a recurrence relation satisfied by
the coefficients; convergence of the series occurs only for a
specific value of the non integer exponent. For any given energy,
two independent solutions of that kind are obtained. They
constitute a basis in the space of solutions of the differential
equation for that value of the energy. The wave function, that can
be written as a linear combination (with only one effective degree
of freedom) of these two basic solutions, must be well behaved at
the origin and at infinity. This double requirement can be
satisfied, by adjusting the degree of freedom of the mentioned
linear combination, only when the considered value of the energy
is one of its eigenvalues. This is, schematically, the procedure
followed by Znojil to find the energies and the wave functions of
certain spiked harmonic oscillators. The procedure, rigorous in
principle, presents the drawback that there is no possibility of
evaluating the basic solutions neither at the origin nor at
infinity. For this reason, Znojil obtained approximate values of
the eigenenergies by requiring the vanishing of the wave function
at a pair of points $r_0\ll 1$ and $r_\infty\gg 1$.

The procedure that we present in this paper is much in the spirit
of the Znojil's method, but differs from it in the manner in which
the regularity of the wave function is imposed: our $r_0$ and
$r_\infty$ are actually the origin and the infinity. Although the
wave function cannot be calculated at these points, we are able to
know its asymptotic behaviour in their vicinity and we may require
that it be the correct one. The eigenenergies determined in this
way are, therefore, exact save for errors inherent to the
computation process. But these errors can be reduced by increasing
the number of digits carried along the calculations.

In the next Section we consider basic solutions of the
Schr\"odinger equation for a very general three-dimensional spiked
oscillator represented by the potential
\begin{equation}
V(r) =  \sum_{q} A(q)\,r^{q},  \label{uno}
\end{equation}
where the index $q$ runs along a finite set of negative and
positive integer and/or rational numbers, whose extremes are
$q_{\mbox{\footnotesize min}}<0$ and $q_{\mbox{\footnotesize
max}}>0$. Moreover, in order to have a true spiked oscillator, we
assume that both $A(q_{\mbox{\footnotesize min}})$ and
$A(q_{\mbox{\footnotesize max}})$ are positive. Section 3 explains
how to determine the behaviour at the origin and at infinity of a
general solution. The requirement, in Section 4, of a regular behaviour at both singular points provides a quantization condition whose fulfilment determines the energy levels and the wavefunction. The procedure is applied to three popular cases of spiked oscillators in Section 5. Appendices A and B give details of the nontrivial steps of the method.

\section{Solutions of the Schr\"odinger equation}

The radial Schr\"odinger equation for the wave of angular momentum $\mathcal{L}$
\begin{equation}
\left[
-\frac{d^2}{dr^2}+\frac{\mathcal{L}(\mathcal{L}+1)}{r^2}+V(r)\right]\,R(r)=E\,R(r),
\label{dos}
\end{equation}
with a potential given by (\ref{uno}), can be easily written in the
form
\begin{equation}
-z^2\frac{d^2w}{dz^2}+g(z)\,w=0,  \label{tres}
\end{equation}
with a ``potential" (including centrifugal and energy terms) that,
multiplied by $z^2$, becomes a combination of positive and
negative integer powers of $z$,
\begin{equation}
g(z)=\sum_{s=-2M}^{2N}g_s\,z^s, \qquad M, N >0, \quad g_{-2M}> 0,
\quad g_{2N}>0. \label{cuatro}
\end{equation}
Obviously, $z$ represents a power of the radial variable
$r$, with exponent conveniently chosen, and $w(z)$ is the reduced radial
wave function $R(r)$ multiplied by an adequate function of $r$.
The origin and the infinity are the only singularities, of ranks\footnote{We adopt
the definition of ranks used in Ref. \cite{naun}, mainly because our work is based
on the Naundorf's treatment of the homogeneous linear second order differential
equation. According to the definitions used in Ref. \cite{slav}, the singularities should
be of ranks $M+1$ and $N+1$.}
respectively $M$ and $N$, of the differential equation
(\ref{tres}). We are interested in considering three pairs of
independent solutions, namely:
\begin{itemize}
\item
Two {\em Floquet} or {\em multiplicative} solutions \cite{slav,arsc}, $w_1$ and $w_2$,
that, except for particular sets of values of the parameters $g_s$
in (\ref{cuatro}), have the form
\begin{equation}
w_j=z^{\nu_j}\sum_{n=-\infty}^{\infty} c_{n,j}\,z^n, \quad
\mbox{being} \; \sum_{n=-\infty}^{\infty} |c_{n,j}|^2<\infty,
\quad j=1,2. \label{cinco}
\end{equation}
The indices $\nu_j$ are not uniquely defined. They admit
addition of any integer (with an adequate relabeling of the
coefficients).
In the general case, the indices $\nu_j$ and the coefficients
$c_{n,j}$ may be complex.
\item
Two {\em Thom\'e}\footnote{Although the credit of these solutions is attributed to Thom\'e, they were used before by Fabry and by Poincar\'e. We are grateful to Prof. Alberto Gr\"unbaum for illustrating us about this fact.} formal solutions, $w_3$ and $w_4$, that have the nature
of asymptotic expansions for $z\to\infty$,
\begin{equation}
\fl w_k(z)\sim\exp\left(\sum_{p=1}^N\frac{\alpha_{p,k}}{p}\,
z^p\right) z^{\mu_k}\,\sum_{m=0}^\infty a_{m,k}\,z^{-m}, \quad
a_{0,k}\neq 0, \quad k=3,4. \label{seis}
\end{equation}
It is usual to say that these two expansions are associated to
each other.
\item
Two {\em Thom\'e} formal solutions, $w_5$ and $w_6$, asymptotic expansions for
$z\to 0$, of the form
\begin{equation}
w_l(z)\sim\exp\left(\sum_{q=1}^M\frac{\beta_{q,l}}{q}\,
z^{-q}\right) z^{\rho_l}\,\sum_{m=0}^\infty b_{m,l}\,z^{m}, \quad
b_{0,l}\neq 0, \quad l=5,6. \label{siete}
\end{equation}
Also these expansions are associated.
\end{itemize}

The determination of the indices $\nu_j$ and the coefficients $c_{n,j}$ of
the multiplicative solutions is rather laborious. By substitution
of (\ref{cinco}) in (\ref{tres}) one obtains the infinite set of
homogeneous equations for the coefficients
\begin{equation}
(n\! +\! \nu_j)(n\! -\! 1\! +\! \nu_j)\,c_{n,j} -
\sum_{s=-2M}^{2N}g_s\,c_{n-s,j}=0,\quad n= \ldots, -1, 0, 1, \ldots,
\label{ocho}
\end{equation}
that can be interpreted as a nonlinear eigenvalue problem, where
the eigenvalue $\nu$ must be such that
\begin{equation}
\sum_{n=-\infty}^{\infty} |c_{n,j}|^2<\infty. \label{nueve}
\end{equation}
In  Appendix A we recall the Newton iterative method to solve that
problem. In general, two indices, $\nu_1$ and $\nu_2$, and two
corresponding sets of coefficients, $\{c_{n,1}\}$ and $\{c_{n,2}\}$
are obtained, but for certain sets of values of the parameters $g_s$
only one multiplicative solution appears. Any other independent
solution must include powers of the variable multiplied by its
logarithm. Such logarithmic solutions cannot correspond, usually, to
the practical system that one tries to describe and are, therefore,
to be discarded. We will assume, from now on, that the parameters
$g_s$ are such that (\ref{tres}) admits two independent
multiplicative solutions.

In what concerns the formal solutions $w_3$ and $w_4$, the
exponents $\alpha$ and $\mu$ and the coefficients $a_m$  must be
such that the expansions in the right hand side of (\ref{seis})
satisfy the differential equation. One obtains in this way
\begin{eqnarray}
\fl\left[\left(\sum_{p=1}^N\alpha_pz^p\right)^2 \! \! +
\sum_{p=2}^N(p\! -\! 1)\alpha_pz^p-g(z)\right]
\sum_{m=0}^{\infty}a_m\, z^{-m} +
2\left(\sum_{p=1}^N\alpha_pz^p\right)
\sum_{m=0}^{\infty}(-m\! +\! \mu)\, a_m\, z^{-m} \nonumber \\
+ \sum_{m=0}^{\infty}
 (-m\! +\! \mu)(-m\! -\! 1\! +\! \mu)\, a_m\, z^{-m} = 0. \label{diez}
\end{eqnarray}
Cancellation of the powers $z^{2N}, z^{2N-1}, \ldots , z^{N+1}$
inside the bracket of the first term in (\ref{diez}) produces a
system of equations
\begin{equation}
\eqalign{
(\alpha_{N})^2 - g_{2N}  =  0,  \\
2\, \alpha_{N}\, \alpha_{N-1} - g_{2N-1}  =  0,    \\
2\, \alpha_{N}\, \alpha_{N-2} + (\alpha_{N-1})^2 - g_{2N-2}  =  0,   \\
  \vdots   \\
2\, \alpha_{N}\, \alpha_{1} + \ldots - g_{N+1} = 0, } \label{duno}
\end{equation}
which can be solved successively. There are two sets of solutions,
$\{\alpha_{p,3}\}$ and $\{\alpha_{p,4}\}$, that obviously verify
\begin{equation}
\alpha_{p,3} =  -\, \alpha_{p,4}, \qquad  p=1, \ldots , N.
\label{ddos}
\end{equation}
Let us assign the labels 3 and 4 in such a way that
\begin{equation}
\alpha_{N,3} = -\sqrt{g_{2N}},  \qquad \alpha_{N,4} =
+\sqrt{g_{2N}}.  \label{dtres}
\end{equation}
Accordingly, the formal solution $w_3$ presents, from the physical
point of view, the adequate behavior at infinity, on the positive real semiaxis,
whereas $w_4$
should be rejected. Cancellation of the coefficient of $z^N$ in the
left hand side of (\ref{diez}) gives the two exponents $\mu_k$ in
terms of the previously obtained $\alpha_{p,k}$. Notice that
\[
\mu_3+\mu_4=-N+1.
\]
Finally, cancellation of the coefficient of
$z^n$, $n=N-1, N-2, \ldots, -\infty$, implies a recurrence relation
for the coefficients $a_{m,k}$ that allows one to obtain all of them
starting with an arbitrarily chosen $a_{0,k}\neq 0$.

Analogously, by
requiring the expansions in the right hand side of
(\ref{siete}) to satisfy the differential equation, one obtains
for the formal solutions $w_5$ and $w_6$
\begin{eqnarray}
\fl\left[\left(\sum_{q=1}^M\beta_qz^{-q}\right)^2 \! \! +
\sum_{q=1}^M(q\! +\! 1)\beta_qz^{-q}-g(z)\right]
\sum_{m=0}^{\infty}b_m\, z^{m} -
2\left(\sum_{q=1}^M\beta_qz^{-q}\right)
\sum_{m=0}^{\infty}(m\! +\! \rho)\, b_m\, z^{m} \nonumber \\
+ \sum_{m=0}^{\infty}
 (m\! +\! \rho)(m\! -\! 1\! +\! \rho)\, b_m\, z^{m} = 0. \label{dcuatro}
\end{eqnarray}
Now, cancellation of the powers $z^{-2M}, z^{-2M+1}, \ldots ,
z^{-M-1}$ inside the bracket of the first term in (\ref{dcuatro})
gives the system of equations
\begin{equation}
\eqalign{
(\beta_{M})^2 - g_{-2M}  =  0,  \\
2\, \beta_{M}\, \beta_{M-1} - g_{-2M+1}  =  0,    \\
2\, \beta_{M}\, \beta_{M-2} + (\beta_{M-1})^2 - g_{-2M+2}  =  0,   \\
  \vdots   \\
2\, \beta_{M}\, \beta_{1} + \ldots - g_{-M-1}  =  0. }
\label{dcinco}
\end{equation}
The two sets of solutions, $\{\beta_{q,5}\}$ and $\{\beta_{q,6}\}$,
verify
\begin{equation}
\beta_{q,5} =  -\, \beta_{q,6}, \qquad  q=1, \ldots , M.
\label{dseis}
\end{equation}
If the labels 5 and 6 are assigned in such a way that
\begin{equation}
\beta_{M,5} = -\sqrt{g_{-2M}},  \qquad \beta_{M,6} =
+\sqrt{g_{-2M}}, \label{dsiete}
\end{equation}
the formal solution $w_5$ vanishes at the origin, whereas $w_6$
diverges. Cancellation of the coefficient of $z^{-M}$ in the left
hand side of (\ref{dcuatro}) allows one to obtain the two exponents
$\rho_l$ in terms of the previously calculated $\beta_{q,l}$. They
obey the relation
\[
\rho_5+\rho_6=M+1.
\]
The coefficients $b_{m,l}$ can be obtained, starting with
an arbitrarily chosen $b_{0,l}$, by making use of the recurrence
relation steming from the cancellation of the coefficient of $z^n$,
$n=-M+1,-M+2, \ldots, +\infty$, in the left hand side of
(\ref{dcuatro}).

\section{The connection factors}

Any solution $w$ of the differential equation (\ref{tres}) can be
written as a linear combination of the two multiplicative
solutions,
\begin{equation}
w=\zeta_1\,w_1+\zeta_2\,w_2.  \label{docho}
\end{equation}
Its behavior in the neighbourhood of the singular points can be
immediately written if, besides the coefficients $\zeta_1$ and
$\zeta_2$, one knows the behaviour of the multiplicative
solutions, that is, if one knows the {\em connection factors} $T$
of their asymptotic expansions,
\begin{eqnarray}
w_j \sim  T_{j,3}\,w_3+T_{j,4}\,w_4, & \qquad \mbox{for}
\; z\to\infty, & \qquad j=1, 2, \label{dnueve} \\
w_j  \sim  T_{j,5}\,w_5+T_{j,6}\,w_6, & \qquad \mbox{for} \; z\to 0,
& \qquad j=1, 2. \label{veinte}
\end{eqnarray}
These connection factors are, obviously, numerical constants, but
their values depend on the sector of the complex plane where $z$
lies. This fact, known as ``Stokes phenomenon" \cite{ding},
introduces a slight complication. As it is well known, the
connection factor multiplying any one of the asymptotic
expansions in the right hand sides of (\ref{dnueve}) and
(\ref{veinte}) takes different values in the sectors of the
complex $z$-plane separated by a Stokes ray of the associated
expansion. On the ray, the value of the connection factor is the
average of those two different ones. We are interested in the
behaviour of $w(z)$ in the vicinity of the origin and at infinity
on the ray $\arg z=0$. Since this is a Stokes ray for the
expansions $w_4$ and $w_6$, the values of $T_{j,3}$ and $T_{j,5}$
for $\arg z=0$ are the respective averages of their values for $z$
just above and below the positive real semiaxis.

The problem of finding the connection factors $T$ has been
considered by several authors. Most of them \cite{kohn} refer to a
differential equation for which the origin is an ordinary or a
regular singular point and the infinity is an irregular singular
one. As far as we know, only the procedure developed by Naundorf
\cite{naun} becomes applicable also to the present case of both the
origin and the infinity being irregular singular points. In a former
paper \cite{gom1} we have suggested a modification of the Naundorf's
procedure, improving notably its performance, and applied it to find
the bound states of anharmonic oscillators, whose Schr\"odinger
equations present an irregular singular point at infinity, the
origin being an ordinary or a regular singular one. Some
mathematical questions left aside in that paper have been tackled in
a posterior one \cite{gom2}. As we are going to show, our
improvement of the Naundorf's method is easily applicable in the
case of the origin being also an irregular singular point, as it
happens in the Schr\"odinger equations of spiked oscillators.

Our procedure to calculate the connection factors makes use of the
fact that they can be written as quotients of Wronskians of the
solutions presented in section 2. We adopt the usual definition of
Wronskian of two functions $f(z)$ and $g(z)$, namely
\[
\mathcal{W}\big[f,g\big](z)=f(z)\,\frac{dg(z)}{dz}-\frac{df(z)}{dz}\,g(z)
\]
We benefit from the fact that equation (\ref{tres}) is a second
order  linear differential one where the first derivative term is
absent and, therefore, the Wronskian of any two of their solutions
is a constant. Then, it is immediate to obtain
\begin{eqnarray}
T_{j,3}  = \frac{\mathcal{W}[w_j,w_4]}{\mathcal{W}[w_3,w_4]}, &
\qquad T_{j,4} = \frac{\mathcal{W}[w_j,w_3]}{\mathcal{W}[w_4,w_3]},
 \qquad j=1,2,   \label{vuno} \\
T_{j,5}  = \frac{\mathcal{W}[w_j,w_6]}{\mathcal{W}[w_5,w_6]}, &
\qquad T_{j,6} = \frac{\mathcal{W}[w_j,w_5]}{\mathcal{W}[w_6,w_5]},
\qquad j=1,2,  \label{vdos}
\end{eqnarray}
The denominators of the quotients in the right hand sides of these
equations can be computed trivially. One obtains
\begin{eqnarray}
\mathcal{W}[w_3,w_4] = - \,\mathcal{W}[w_4,w_3] =
-\, 2\alpha_{N,3}\,a_{0,3}\,a_{0,4},  \label{vtres}  \\
\mathcal{W}[w_5,w_6] = - \,\mathcal{W}[w_6,w_5] =
2\beta_{M,5}\,b_{0,5}\,b_{0,6}.  \label{vcuatro}
\end{eqnarray}
The calculation of the numerators is not so easy. Direct computation
of any one of them gives a certain power of $z$ times an infinite
series of negative and positive powers of $z$. This does not seem to
be an adequate expression to determine the (independent of $z$)
value of the Wronskian. Instead, we have devised a trick that allows
one to obtain each one of those numerators. Appendix B contains a
detailed description of the procedure.

\section{The quantization condition}

We have already mentioned that any solution of (\ref{tres}) can be
written as a linear combination of the two multiplicative solutions,
like in (\ref{docho}). According to (\ref{veinte}), its behaviour in
the neighbourhood of the origin is given by
\begin{equation}
\fl w(z)\sim(\zeta_1\,T_{1,5}+\zeta_2\,T_{2,5})w_5(z) +
(\zeta_1\,T_{1,6}+\zeta_2\,T_{2,6})w_6(z) \qquad \mbox{for}\;z\to 0.
\label{vcinco}
\end{equation}
Since $w_6$ diverges as $z$ goes to zero, the coefficients $\zeta_1$
and $\zeta_2$ must be such that
\begin{equation}
T_{1,6}\,\zeta_1+T_{2,6}\,\zeta_2 = 0. \label{vseis}
\end{equation}
On the other hand, at infinity,
\begin{equation}
\fl w(z)\sim(\zeta_1\,T_{1,3}+\zeta_2\,T_{2,3})w_3(z) +
(\zeta_1\,T_{1,4}+\zeta_2\,T_{2,4})w_4(z) \qquad \mbox{for}\;z\to
\infty. \label{vsiete}
\end{equation}
The correct asymptotic behaviour occurs when
\begin{equation}
T_{1,4}\,\zeta_1+T_{2,4}\,\zeta_2 = 0. \label{vocho}
\end{equation}
The fulfilment of both requirements (\ref{vseis}) and (\ref{vocho})
implies the quantization condition
\begin{equation}
T_{1,6}T_{2,4} - T_{1,4}T_{2,6} = 0. \label{vnueve}
\end{equation}
For given potential (\ref{uno}) and angular momentum, the connection
factors are functions only of the energy, through one of the
parameters $g_s$ in (\ref{cuatro}). Therefore, the left hand side of
(\ref{vnueve}) is but a function of the energy whose zeros are the
eigenenergies of the spiked oscillator.

The problem of finding the wave functions is now immediately solved.
Any nontrivial solution $\{\hat{\zeta_1}, \hat{\zeta_2}\}$ of the
system of equations (\ref{vseis}) and (\ref{vocho}) gives the
(unnormalized) physical solution
\begin{equation}
w_{\rm phys}(z)= \hat{\zeta_1}\,w_1(z) +\hat{\zeta_2}\,w_2(z).
\label{treinta}
\end{equation}
For the computation of $w_{\rm phys}$ in the neighbourhood of the
origin or for large values of $z$, the asymptotic expansions
\begin{eqnarray}
w_{\rm phys}(z) \sim (\hat{\zeta_1}\,T_{1,5}
+\hat{\zeta_2}\,T_{2,5})w_5(z),  \qquad z \to 0, \label{tuno} \\
w_{\rm phys}(z) \sim (\hat{\zeta_1}\,T_{1,3}
+\hat{\zeta_2}\,T_{2,3})w_3(z),  \qquad z \to \infty, \label{tdos}
\end{eqnarray}
should be used instead of (\ref{treinta}), that suffers from strong
cancellations in the right hand side for $z$ in those regions. So,
there is no difficulty to normalize $w_{\rm phys}$ properly.

\section{Some examples}

In order to illustrate how our procedure may serve to treat spiked oscillators, we are going to apply it to
a few cases that have already been considered, by having recourse to different approximations,
 by other authors. The numerical results given
below have been obtained by using double precision FORTRAN codes. In our opinion, the digits shown
in the quoted values of the energies are correct. Our procedure could provide additional correct digits
but it would require a higher precision arithmetic. This need is more evident in the case of high values
of the potential parameters, angular momentum different from zero, or excited states.

Following a common practice and to facilitate comparison of our results with those of other authors,
we assume that $r$ is a dimensionless variable that represents a distance in a given scale. Analogously, we
use dimensionless symbols, $E$ and $A_q$ to represent the energy and
the intensities of the different terms of the potential, that are assumed expressed in adequate units.

\subsection{Potential $V(r)=A_2\,r^2+A_{-4}\,r^{-4}$}

The first example to be considered is a three dimensional spiked harmonic oscillator, of potential
\begin{equation}
 V(r)=A_2\,r^2+A_{-4}\,r^{-4},   \label{ttres}
\end{equation}
which has been most discussed by other authors. We will assume, without loss of generality, that
\[
A_2=1\,.
\]
The radial Schr\"odinger equation (\ref{dos}), written in terms of the variable
\[
z\equiv r\,,
\]
turns into Eq. (\ref{tres}) for the function
\[
w(z)=R(r),
\]
with
\[
g(z)=A_{-4}\,z^{-2} + \mathcal{L}(\mathcal{L}+1) - E\,z^2 + z^4,
\]
where, obviously, $\mathcal{L}$ represents the angular momentum quantum number.
So, the ranks of the singularities at the origin and at infinity are, respectively,
\[
M=1, \qquad N=2.
\]
The coefficients $c_{n,j}$ of the Floquet solutions obey the recurrence relation
(omitting the second subindex $j$)
\[
A_{-4}\,c_{n+6}+\left[\mathcal{L}(\mathcal{L}+1)-(n+4+\nu)(n+3+\nu)\right]\,c_{n+4}-E\,c_{n+2}+c_n=0.
\]
The Thom\'e solutions at infinity have exponents
\[
\alpha_{2,3}=-\alpha_{2,4}=-1,\quad \alpha_{1,3}=-\alpha_{1,4}=0,
\quad \mu_3=\frac{-1+E}{2}\quad \mu_4=\frac{-1-E}{2},
\]
and coefficients $a_{m,k}$ ($k=3, 4$) obeying the recurrence relation
(omitting the second subindex $k$)
\[
2\,\alpha_2\,m\,a_m=\left[(m-2-\mu)(m-1-\mu)-\mathcal{L}(\mathcal{L}+1)\right]\,a_{m-2}-A_{-4}\,a_{m-4}.
\]
For the Thom\'e solutions at the origin the exponents are
\[
\beta_{1,5}=-\beta_{1,6}=-A_{-4}^{1/2}, \qquad \rho_5=\rho_6=1,
\]
and the coefficients $b_{m,k}$, ($k=5, 6$) satisfy the recurrence relation
(omitting the second subindex $k$)
\[
2\,\beta_1\,m\,b_m=\left[m(m-1)-\mathcal{L}(\mathcal{L}+1)\right]\,b_{m-1}+E\,b_{m-3}-b_{m-5}.
\]
The definitions of Floquet and Thom\'e solutions given above fix them up to
arbitrary multiplicative constants. To avoid ambiguities in the definition of the
connection factors, we assume from now on that those arbitrary constants have been
chosen in such a way that
\[
c_{0,1}=c_{0,2}=1,\quad a_{0,3}=a_{0,4}=1, \quad b_{0,5}=b_{0,6}=1.
\]
Then, according to Eqs. (\ref{vtres}) and (\ref{vcuatro}), we have for the denominators
in Eqs. (\ref{vuno}) and (\ref{vdos})
\begin{eqnarray}
\mathcal{W}[w_3,w_4] = - \,\mathcal{W}[w_4,w_3] = 2,  \label{tcuatro}  \\
\mathcal{W}[w_5,w_6] = - \,\mathcal{W}[w_6,w_5] = -A_{-4}^{1/2}.  \label{tcinco}
\end{eqnarray}
Our procedure gives for the numerators in the same equations (for $j=1, 2$)
\begin{eqnarray}
\mathcal{W}[w_j,w_3] = 2^{n+\delta_1^{(j,3)}}\,\Gamma (n+1+\delta_1^{(j,3)})\, \gamma_{2n+1}^{(j,3)},  \label{tseis}\\
\mathcal{W}[w_j,w_4] =(-1)^n\,\cos(\pi\delta_1^{(j,4)})\, 2^{n+\delta_1^{(j,4)}}\,\Gamma (n+1+\delta_1^{(j,4)})\, \gamma_{2n+1}^{(j,4)},
\label{tsiete} \\
\mathcal{W}[w_j,w_5] =A_{-4}^{(-n+\nu_j+1)/2}\,\Gamma (n-\nu_j)\,\gamma_{-n}^{(j,5)}, \label{tocho} \\
\mathcal{W}[w_j,w_6] =(-1)^{n-1}\,\cos(\pi\nu_j)\,A_{-4}^{(-n+\nu_j+1)/2}\,\Gamma (n-\nu_j)\,\gamma_{-n}^{(j,6)}, \label{tnueve}
\end{eqnarray}
where we have abbreviated
\begin{eqnarray}
\fl\delta_1^{(j,k)}=(\nu_j+\mu_k+1)/2\,, \qquad k=3, 4,  \label{cuarenta} \\
\fl\gamma_m^{(j,k)}=\sum_{s=0}^\infty a_{s,k}\left(\alpha_{2,k}\,c_{m+s-1,j}-(m\! +\! 2s\! +\! 1\! +\! \nu_j\! -\! \mu_k)
\,c_{m+s+1,j}\right),\qquad k=3, 4, \label{cuno} \\
\fl\gamma_m^{(j,l)}=\sum_{s=0}^\infty b_{s,l}\left(-\ \beta_{1,l}\,c_{m-s+2,j}+(2s-m-\nu_j)\,c_{m-s+1,j}\right), \qquad l=5, 6.  \label{cdos}
\end{eqnarray}

We report in Table 1 our ground-state energies of the spiked oscillator of potential (\ref{ttres}),
for several values of the parameter $A_{-4}$. The quoted energies should be compared
with those in Table 3 of Ref. \cite{hall3}, that collects results obtained by several authors with different methods.
Obviously, our procedure is superior to variational ones, as it provides with four more correct digits in the energies for
the values of $A_{-4}$ considered. In fact, our double precision FORTRAN calculations give results with an accuracy comparable to the excellent one reached by Buend\'{\i}a {\it et al.} \cite{buen} thanks to an analytic continuation method or by Roy \cite{roy} by using a generalized pseudospectral method. (Notice that, due to a different definition of the Schr\"odinger operator, our energies should be divided by 2 if they are to be compared with those of Table 4 of Ref. \cite{roy}.) As mentioned above, larger values of $A_{-4}$ require a more precise arithmetic to
the same accuracy in the results. For completeness, we give also in the same table the indices $\nu_j$ of the Floquet
solutions. We have already mentioned the existing ambiguity in the values of the indices. To avoid it, we assume that the indices vary continuously with the parameter $A_{-4}$ and fix their integer part in such a way that $\nu_1=0$ (and, consequently, $\nu_2=1$) for $A_{-4}=0$, as it should
be for a particle of zero angular momentum. As $A_{-4}$ increases, the index $\nu_1$ increases also until it reaches the value
0.5 for $A_{-4}= 0.1305\ldots$ and an eigenenergy $E=3.6454\ldots$. Notice that, for those critical values of the parameters,
$\nu_1=\nu_2$ and only one Floquet solution of the form (\ref{cinco}) is obtained. Any other independent solution
of the differential equation must contain logarithmic terms. Our procedure, in its present form, is not applicable in
this case. A different implementation of the basic idea would be necessary. If $A_{-4}$ continues increasing
above the critical value, the real part of $\nu_1$ remains equal to $0.5$, whereas its imaginary part increases. Obviously,
the two Floquet solutions are then complex conjugate to each other. This fact allows one to simplify the calculations in our procedure.
\begin{table}
\caption{Ground-state energy of a spiked oscillator of potential (\ref{ttres}) for several values of the intensity $A_{-4}$ of the singular repulsive potential. The index $\nu_1$ of one of the Floquet solutions, $w_1$, is also given. The index $\nu_2$ of the other Floquet solution, $w_2$, is immediately obtained from the relation $\nu_2=1-\nu_1$.}
\begin{indented}
\item[]
\begin{tabular}{lll}
\br
 $A_{-4}$ & $\qquad E$ & $\qquad \nu_1$ \\
\mr
0.0001 & 3.0222745087 & 0.201485000573E$-$03 \\
0.001 & 3.0687631709 & 0.204586237797E$-$02 \\
0.005 & 3.1483523083 & 0.104967473634E$-$01 \\
0.01 & 3.2050674951 & 0.213850813448E$-$01 \\
0.1 & 3.5755519912 & 0.270240464647E+00 \\
0.4 & 4.0319714400 & 0.5 + $i\,$0.606083134346E+00 \\
1 & 4.4941779834 & 0.5 + $i\,$0.950268234562E+00 \\
10 & 6.6066225120 & 0.5 + $i\,$0.203793867918E+01 \\
100 & 11.265080432 & 0.5 + $i\,$0.412681646514E+01 \\
\br
\end{tabular}
\end{indented}
\end{table}

In order to illustrate the trend of the coefficients of the Floquet and Thom\'e solutions, we show, in
Tables 2 to 7, the most relevant of them for two particular cases. Tables 2, 3 and 4 correspond to the ground
state of the example discussed by Znojil \cite{znoj3}, namely that of $A_{-4}=0.4$. In this case,
the energy and the indices of the Floquet solutions are
\[
E=4.0319714400\,, \qquad \nu_1=\overline{\nu_2}=0.5+i\,0.606083134346,
\]
and the connection factors
\begin{eqnarray}
T_{1,3}=\overline{T_{2,3}}= 0.363722440835 + i\,1.049440656062\,,    \nonumber \\
T_{1,4}=\overline{T_{2,4}}= -\,0.378572197756 + i\,0.137728550255\,,    \nonumber \\
T_{1,5}=\overline{T_{2,5}}= 0.520935174155 + i\,1.431885933657\,,   \nonumber  \\
T_{1,6}=\overline{T_{2,6}}= 0.436272922113 - i\,0.158720681109\,.    \nonumber
\end{eqnarray}
The coefficients of the Floquet solutions can be seen in Table 2, and those of the Thom\'e
solutions at infinity and at the origin in Tables 3 and 4, respectively. The second particular case, illustrated in Tables 5, 6 and 7, is that of
$A_{-4}=1$ and angular momentum $\mathcal{L}=2$, considered by Aguilera-Navarro and Ley Koo \cite[Table II]{agui3}. For the energy and the indices of the Floquet solutions we have obtained
\[
\fl E=7.2242871639\,, \qquad \nu_1=-2.083592228877\,, \qquad \nu_2=3.083592228877\,,
\]
and for the connection factors
\begin{eqnarray}
T_{1,3}=-\,0.7041938314\,, \qquad   & T_{2,3}= 0.9280831701\,,  \nonumber \\
T_{1,4}=-\,2.4617003408\,, \qquad   & T_{2,4}= -\,0.0462330978\,,   \nonumber \\
T_{1,5}=-\,7.7185228435\,, \qquad    & T_{2,5}= 0.1449612757\,,    \nonumber  \\
T_{1,6}=7.9925473468\,, \qquad    & T_{2,6}= 0.1501077191\,.    \nonumber
\end{eqnarray}
The ambiguity in the integer part of the indices of the Floquet solutions has been resolved in this second particular case bearing in mind that, since $\mathcal{L}=2$, it is desirable that $\nu_1\to -2$ and $\nu_2\to 3$ for $A_{-4}\to 0$.
\begin{table}
\caption{Some coefficients of the Floquet solution $w_1$, normalized in such a way that $c_{0,1}=1$, for the particular case of $A_{-4}=0.4$, $\mathcal{L}=0$ and energy $E=4.0319714400$. The coefficients with odd first label are equal to zero. The other Floquet solution, $w_2$, has coefficients $c_{m,2}$ complex conjugate of the $c_{m,1}$.}
\begin{indented}
\item[]
\begin{tabular}{rrr}
\br
 $n$ &  $\Re c_{2n,1}\qquad$ & $\Im c_{2n,1}\qquad$  \\
\mr
$-$10 & $-$0.142317651396E$-$23 & 0.690261395778E$-$23 \\
$-$9 & $-$0.100263927421E$-$20 & 0.697827393080E$-$20 \\
$-$8 & $-$0.429957466419E$-$18 & 0.569639303294E$-$17 \\
$-$7 & 0.167970940662E$-$17 & 0.365781757971E$-$14 \\
$-$6 & 0.156002041860E$-$12 & 0.178667148961E$-$11 \\
$-$5 & 0.120892275974E$-$09 & 0.634808135173E$-$09 \\
$-$4 & 0.492753885758E$-$07 & 0.154076769690E$-$06 \\
$-$3 & 0.115445648553E$-$04 & 0.232262850896E$-$04 \\
$-$2 & 0.144400180631E$-$02 & 0.184616207302E$-$02 \\
$-$1 & 0.780262828557E$-$01 & 0.537272549009E$-$01 \\
0 & 1\phantom{.000000000000E+00} &          \\
1 & $-$0.760452558754E+00 & 0.536951488093E+00 \\
2 & 0.199603204168E+00 & $-$0.202477697179E+00 \\
3 & $-$0.348689735973E$-$01 & 0.453398033971E$-$01 \\
4 & 0.433386333626E$-$02 & $-$0.673686840525E$-$02 \\
5 & $-$0.431181023059E$-$03 & 0.781829457162E$-$03 \\
6 & 0.349885697033E$-$04 & $-$0.725041819983E$-$04 \\
7 & $-$0.243278698241E$-$05 & 0.570826186141E$-$05 \\
8 & 0.146155963833E$-$06 & $-$0.385089511701E$-$06 \\
9 & $-$0.779445127491E$-$08 & 0.229774575047E$-$07 \\
10 & 0.370654869957E$-$09 & $-$0.121862184787E$-$08 \\
\br
\end{tabular}
\end{indented}
\end{table}
\begin{table}
\caption{The first coefficients of the two Thom\'{e} solutions at infinity for the particular case of $A_{-4}=0.4$, $\mathcal{L}=0$ and energy $E=4.0319714400$. We have skipped over the coefficients with odd first label, since they are equal to zero.}
\begin{indented}
\item[]
\begin{tabular}{rrr}
\br
 $m$ & $a_{2m,3}\qquad$ & $a_{2m,4}\qquad$ \\
\mr
0 & 1\phantom{.000000000000E+00} & 1\phantom{.000000000000E+00} \\
1 & $-$0.195556745811E+00 & 0.221154246581E+01 \\
2 & 0.675581627248E$-$01 & 0.683622152000E+01 \\
3 & $-$0.552411742467E$-$01 & 0.278260729988E+02 \\
4 & 0.865891090477E$-$01 & 0.140764673339E+03 \\
5 & $-$0.212297950675E+00 & 0.851787239516E+03 \\
6 & 0.709506391974E+00 & 0.6001543154448E+04 \\
7 & $-$0.305385835605E+01 & 0.482637790136E+05 \\
8 & 0.160735596654E+02 & 0.436250311536E+06 \\
9 & $-$0.100168048970E+03 & 0.437841970329E+07 \\
10 & 0.721888409401E+03 & 0.483138693754E+08 \\
\br
\end{tabular}
\end{indented}
\end{table}
\begin{table}
\caption{The first coefficients of the two Thom\'{e} solutions at the origin for the particular case of $A_{-4}=0.4$, $\mathcal{L}=0$ and energy $E=4.0319714400$.}
\begin{indented}
\item[]
\begin{tabular}{rrr}
\br
 $m$ & $b_{m,5}\qquad$ & $b_{m,6}\qquad$ \\
\mr
0 & 1\phantom{.000000000000E+00} & 1\phantom{.000000000000E+00} \\
1 & 0\phantom{.000000000000E+00} & 0\phantom{.000000000000E+00} \\
2 & 0\phantom{.000000000000E+00} & 0\phantom{.000000000000E+00} \\
3 & $-$0.106251776760E+01 & 0.106251776760E+01 \\
4 & 0.251998215001E+01 & 0.251998215001E+01 \\
5 & $-$0.781076937396E+01 & 0.781076937396E+01 \\
6 & 0.314392488782E+02 & 0.314392488782E+02 \\
7 & $-$0.150276962631E+03 & 0.150276962631E+03 \\
8 & 0.834637749300E+03 & 0.834637749300E+03 \\
9 & $-$0.528962618019E+04 & 0.528962618019E+04 \\
10 & 0.376836341629E+05 & 0.376836341629E+05 \\
\br
\end{tabular}
\end{indented}
\end{table}
\begin{table}
\caption{Coefficients of the Floquet solutions $w_1$ and $w_2$, normalized in such a way that $c_{0,1}=c_{0,2}=1$, for the particular case of $A_{-4}=1$, $\mathcal{L}=2$ and energy $E=7.2242871639$. The coefficients with odd first label are equal to zero.}
\begin{indented}
\item[]
\begin{tabular}{rrr}
\br
 $n$ & $c_{2n,1}\qquad$ & $c_{2n,2}\qquad$ \\
\mr
$-$10 & 0.418079159329E$-$21 & 0.508287996884E$-$16   \\
$-$9 & 0.210620217409E$-$18 & 0.151012739304E$-$13 \\
$-$8 & 0.879231161250E$-$16 & 0.349503786349E$-$11 \\
$-$7 & 0.298162410599E$-$13 & 0.607371865088E$-$09 \\
$-$6 & 0.801421283847E$-$11 & 0.753905871024E$-$07 \\
$-$5 & 0.165459858432E$-$08 & 0.621798117715E$-$05 \\
$-$4 & 0.251717006169E$-$06 & 0.303689920293E$-$03 \\
$-$3 & 0.266341461331E$-$04 & 0.705627796961E$-$02 \\
$-$2 & 0.179770550637E$-$02 & 0.404457498845E$-$01 \\
$-$1 & 0.668756857204E$-$01 & $-$0.120970341064E+00 \\
0 & 1\phantom{.000000000000E+00} & 1\phantom{.000000000000E+00} \\
1 & 0.906280257537E+00 & $-$0.489421426933E+00 \\
2 & 0.180182069867E+01 & 0.121731165609E+00 \\
3 & $-$0.209931807620E+01 & $-$0.202632214005E$-$01 \\
4 & 0.728418847644E+00 & 0.253270584477E$-$02 \\
5 & $-$0.150546747680E+00 & $-$0.253385266560E$-$03 \\
6 & 0.220037264283E$-$01 & 0.211291907657E$-$04 \\
7 & $-$0.249248011094E$-$02 & $-$0.151036531706E$-$05 \\
8 & 0.230170752863E$-$03 & 0.944714256302E$-$07 \\
9 & $-$0.179507697993E$-$04 & $-$0.525254239416E$-$08 \\
10 & 0.121105001156E$-$05 & 0.262829669113E$-$09 \\
\br
\end{tabular}
\end{indented}
\end{table}
\begin{table}
\caption{The first coefficients of the two Thom\'{e} solutions at infinity for the particular case of $A_{-4}=1$, $\mathcal{L}=2$ and energy $E=7.2242871639$. We have skipped over the coefficients with odd first label, since they are equal to zero.}
\begin{indented}
\item[]
\begin{tabular}{rrr}
\br
 $m$ & $a_{2m,3}\qquad$ & $a_{2m,4}\qquad$  \\
\mr
0 & 1\phantom{.000000000000E+00} & 1\phantom{.000000000000E+00} \\
1 & $-$0.143323523200E+00 & 0.375546710516E+01 \\
2 & 0.197417671158E$-$01 & 0.174648768948E+02 \\
3 & $-$0.483024944799E$-$02 & 0.985368170788E+02 \\
4 & 0.281201497378E$-$02 & 0.653978688776E+03 \\
5 & $-$0.344425647424E$-$02 & 0.499199894361E+04 \\
6 & 0.705309939217E$-$02 & 0.430838076974E+05 \\
7 & $-$0.207487524900E$-$01 & 0.414831591842E+06 \\
8 & 0.802543608266E$-$01 & 0.440833793663E+07 \\
9 & $-$0.386209034478E+00 & 0.512489320365E+08 \\
10 & 0.222788471661E+01 & 0.646983198568E+09 \\
\br
\end{tabular}
\end{indented}
\end{table}
\begin{table}
\caption{The first coefficients of the two Thom\'{e} solutions at the origin for the particular case of $A_{-4}=1$, $\mathcal{L}=2$ and energy $E=7.2242871639$.}
\begin{indented}
\item[]
\begin{tabular}{rrr}
\br
 $m$ & $b_{m,5}\qquad$ & $b_{m,6}\qquad$ \\
\mr
0 & 1\phantom{.000000000000E+00} & 1\phantom{.000000000000E+00} \\
1 & 3\phantom{.000000000000E+00} & $-$3\phantom{.000000000000E+00} \\
2 & 3\phantom{.000000000000E+00} & 3\phantom{.000000000000E+00} \\
3 & $-$0.120404786066E+01 & 0.120404786066E+01 \\
4 & $-$0.180607179098E+01 & $-$0.180607179098E+01 \\
5 & 0.461214358197E+00 & $-$0.461214358197E+00 \\
6 & 0.524369089808E$-$01 & 0.524369089808E$-$01 \\
7 & 0.101141803810E+01 & $-$0.101141803810E+01 \\
8 & $-$0.344418092084E+01 & $-$0.344418092084E+01 \\
9 & 0.125072805387E+02 & $-$0.125072805387E+02 \\
10 & $-$0.528728562620E+02 & $-$0.528728562620E+02 \\
\br
\end{tabular}
\end{indented}
\end{table}

Once the connection factors are known, the wave function is easily obtained following the procedure indicated in Section 4.

\subsection{Potential $V(r)=A_2\,r^2+A_{-4}\,r^{-4}+A_{-6}\,r^{-6}$}

Our second example is a three dimensional spiked harmonic oscillator which has received also considerable attention and which becomes quasi-exactly solvable for certain sets of parameters. The potential is
\begin{equation}
 V(r)=A_2\,r^2+A_{-4}\,r^{-4}+A_{-6}\,r^{-6}.   \label{ctres}
\end{equation}
Following a common practice, we will assume
\[
A_2=1\,.
\]
We can reduce the rank of the singularities of the radial Schr\"odinger equation by using a variable
\[
z\equiv r^2\,.
\]
Equation (\ref{dos}) turns in this way into Eq. (\ref{tres}) for the function
\[
w(z) = r^{1/2}\,R(r),
\]
with
\[
g(z)=\frac{1}{4}\left(A_{-6}\,z^{-2}+A_{-4}\,z^{-1} + \mathcal{L}(\mathcal{L}+1)-3/4 - E\,z + z^2\right).
\]
Now the ranks of the singularities at the origin and at infinity are, respectively,
\[
M=1, \qquad N=1.
\]
This fact reveals that we are dealing with a double confluent Heun Equation \cite{schw}. In a recent paper \cite{abad}, we have
detailed the algorithm resulting from the application of our method to an equation of this type.
For the coefficients $c_{n,j}$ of the Floquet solutions we have the recurrence relation
(omitting the second subindex $j$)
\[
\fl A_{-6}\,c_{n+4}+A_{-4}\,c_{n+3}+\left[\mathcal{L}(\mathcal{L}+1)-3/4-4(n+2+\nu)(n+1+\nu)\right]\,c_{n+2}-E\,c_{n+1}+c_n=0.
\]
The Thom\'e solutions at infinity have exponents
\[
\alpha_{1,3}=-\alpha_{1,4}=-1/2\,,
\qquad \mu_3=E/4\,, \quad \mu_4=-E/4\,,
\]
and the recurrence relation for the coefficients $a_{m,k}$ ($k=3, 4$) is
(omitting the second subindex $k$)
\[
\fl 8\,\alpha_1\,m\,a_m=\left[4(m-\mu)(m-1-\mu)-\mathcal{L}(\mathcal{L}+1)+3/4\right]\,a_{m-1}-A_{-4}\,a_{m-2}-A_{-6}\,a_{m-3}.
\]
For the Thom\'e solutions at the origin, the exponents are in this second example
\[
\fl \beta_{1,5}=-\beta_{1,6}=-A_{-6}^{1/2}/2\,, \qquad
\rho_5=1+A_{-4}/4A_{-6}^{1/2}\,, \quad \rho_6=1-A_{-4}/4A_{-6}^{1/2}\,,
\]
and the coefficients $b_{m,l}$ ($l=5, 6$) obey the recurrence relation
(omitting the second subindex $l$)
\[
\fl 8\,\beta_1\,m\,b_m=\left[4(m-1+\rho)(m-2+\rho)-\mathcal{L}(\mathcal{L}+1)+3/4\right]\,b_{m-1}+E\,b_{m-2}-b_{m-3}.
\]
By choosing, as in the first example,
\[
c_{0,1}=c_{0,2}=1,\quad a_{0,3}=a_{0,4}=1, \quad b_{0,5}=b_{0,6}=1.
\]
we have, from Eqs. (\ref{vtres}) and (\ref{vcuatro}),
\begin{eqnarray}
\mathcal{W}[w_3,w_4] = - \,\mathcal{W}[w_4,w_3] = 1,  \label{ccuatro}  \\
\mathcal{W}[w_5,w_6] = - \,\mathcal{W}[w_6,w_5] = -A_{-6}^{1/2}.  \label{ccinco}
\end{eqnarray}
For the numerators in Eqs. (\ref{vuno}) and (\ref{vdos}) our procedure gives in this second example (for $j=1, 2$)
\begin{eqnarray}
\mathcal{W}[w_j,w_3] = 2^{n+\delta^{(j,3)}}\,\Gamma (n+1+\delta^{(j,3)})\, \gamma_{n}^{(j,3)},  \label{cseis}\\
\mathcal{W}[w_j,w_4] =(-1)^n\,\cos(\pi\delta^{(j,4)}) \,2^{n+\delta^{(j,4)}}\,\Gamma (n+1+\delta^{(j,4)})\, \gamma_{n}^{(j,4)},
\label{csiete} \\
\mathcal{W}[w_j,w_5] =(A_{-6}^{1/2}/2)^{-n-\delta^{(j,5)}}\,\Gamma (n+1+\delta^{(j,5)})\,\gamma_{-n}^{(j,5)}, \label{cocho} \\
\mathcal{W}[w_j,w_6] =(-1)^n\,\cos(\pi\delta^{(j,6)})\,(A_{-6}^{1/2}/2)^{-n-\delta^{(j,6)}}\,\Gamma (n+1+\delta^{(j,6)})\,\gamma_{-n}^{(j,6)}, \label{cnueve}
\end{eqnarray}
with the abbreviations
\begin{eqnarray}
\fl\delta^{(j,k)}=\nu_j+\mu_k\,, \qquad k=3, 4,   \qquad\qquad
\delta^{(j,l)}=-\nu_j-\rho_l\,, \qquad l=5, 6,   \label{cincuenta} \\
\fl\gamma_m^{(j,k)}=\sum_{s=0}^\infty a_{s,k}\left(\alpha_{1,k}\,c_{m+s,j}-(m\! +\! 2s\! +\! 1\! +\! \nu_j\! -\! \mu_k)
\,c_{m+s+1,j}\right),\qquad k=3, 4,  \label{quno} \\
\fl\gamma_m^{(j,l)}=\sum_{s=0}^\infty b_{s,l}\left(-\ \beta_{1,l}\,c_{m-s+2,j}+(2s\! -\! m\! -\! 1\! -\! \nu_j\! +\! \rho_l)\,c_{m-s+1,j}\right), \qquad l=5, 6.  \label{qdos}
\end{eqnarray}
The connection factors are then obtained immediately from Eqs. (\ref{vuno}) and (\ref{vdos}).

The ground state energies of the spiked oscillator of potential (\ref{ctres}) obtained with our method, for several values of $A_{-6}$ and $A_{-4}$, are shown in Table 8. The chosen values of the parameters allow to compare our results with the extremely precise ones of Buend\'{\i}a {\it et al.} \cite{buen} and of Roy \cite{roy}. The concordance is remarkable. We have considered also values of  $A_{-6}$ and $A_{-4}$, both different from zero, allowing comparison with the variational results of Saad, Hall and Katatbeth \cite{saad}. In the same table we give also the indices $\nu_1$ and $\nu_2$ of the Floquet solutions. As in the preceding example, the physical solution $w_{\scriptstyle{\rm phys}}$ can be obtained easily following the steps indicated in Section 4. Going back to the original variable $r$ and reduced radial wave function $R(r)$ is trivial.
\begin{table}
\caption{Ground-state energy of a spiked oscillator of potential (\ref{ctres}) for several values of the intensities $A_{-6}$ and $A_{-4}$. The column headed by $E$ contains the results obtained with our method. For comparison, we show, in the column headed by $E_{\scriptstyle{\rm lit}}$, three set of values taken from the literature, namely from Refs. \cite{buen}, \cite{roy}  and \cite{saad}, respectively. (Notice that, due to a different definition of the Hamiltonian, a factor 2 has been applied to the results of Ref. \cite{roy} and, consequently, the last digit may oscillate by one unit.) The last column shows the indices $\nu_1$ and $\nu_2$ of the Floquet solutions.}
\begin{indented}
\item[]
\begin{tabular}{lllll}
\br
 $A_{-6}$ & $A_{-4}$ & $\qquad E$ & $\qquad E_{\scriptstyle{\rm lit}}$ & $\qquad \nu_1=-\nu_2$ \\
\mr
0.001 & 0 & 3.27985582592 & 3.279855825921856 & 0.249216175554 \\
0.0025 & 0 & 3.35391931711 & 3.353919317108725 & 0.247958538878 \\
0.01 & 0 & 3.50545227600 & 3.505452275995097 & 0.241137578178 \\
1 & 0 & 4.65993996957 & 4.659939969573538 & $i\,$0.466911061788 \\
10 & 0 & 6.00320902890 & 6.003209028895745 & $i\,$0.895534935089 \\
\mr
0.005 & 0 &  3.42288418426 & 3.42288418426 & 0.245761020193   \\
0.05 & 0 &  3.76554020606  & 3.76554020604  & 0.198535942381  \\
0.5  & 0 &  4.38790906027  &  4.38790906026  & $i\,$0.337261268644  \\
5  & 0 &  5.51315901419  & 5.51315901418  &  $i\,$0.768433078693   \\
\mr
1 & 10 & 6.67905366445  & 6.679054 & 0.5$\,-\,i\,$1.00539309301  \\
10 & 1 & 6.14012287178 & 6.140123 & $i\,$0.896525791611 \\
10 & 10 & 7.13826093998 & 7.138261 & 0.5$\,-\,i\,$0.320864634688 \\
\br
\end{tabular}
\end{indented}
\end{table}

\subsection{Potential $V(r)=r^2+\lambda \,r^{-5/2}$}

As a third example, we have chosen the spiked oscillator whose potential
\begin{equation}
V(r)=r^2+\lambda\, r^{-5/2}    \label{qtres}
\end{equation}
presents a critical singularity. The Schr\"odinger equation (\ref{dos}) adopts the form (\ref{tres}) for the variables
\[
z \equiv r^{1/4} \qquad \mbox{and} \qquad w(z)= r^{-3/8}R(r),
\]
$g(z)$ being now
\[
g(z)=16\left(\lambda\,z^{-2} + \mathcal{L}(\mathcal{L}+1)+15/64 - E\,z^8 + z^{16}\right).
\]
The singularities at the origin and at infinity have ranks
\[
M=1, \qquad\qquad N=8.
\]
The recurrence relation for the coefficients $c_{n,j}$ of the Floquet solutions (\ref{cinco}) is (omitting the subindex $j=1, 2$)
\[
\fl\lambda c_{n+18}+\left[\mathcal{L}(\mathcal{L}+1)+15/64-(n+16+\nu)(n+15+\nu)/16\right]c_{n+16}-Ec_{n+8}+c_n =0.
\]
For the exponents of the Thom\'e solutions at infinity we have
\[
\fl\alpha_{8,3}=-\alpha_{8,4}=-4\,, \quad \alpha_{7,j}=\alpha_{6,j}=\ldots =\alpha_{1,j}=0\,,\quad \mu_3=-7/2+2E,\quad \mu_4=-7/2-2E,
\]
and for their coefficients $a_{m,k}$ (omitting the subindex $k=3, 4$)
\[
\fl 2\,\alpha_8\,m\,a_m=\left[(m-8-\mu)(m-7-\mu)-16\,\mathcal{L}(\mathcal{L}+1)-15/4\right]\,a_{m-8}-16\lambda\,a_{m-10}.
\]
The Thom\'e solutions at the origin have exponents
\[
\beta_{1,5}=-\beta_{1,6}=-4\,\lambda^{1/2}, \qquad \rho_5=\rho_6=1,
\]
and coefficients $b_{m,l}$ that satisfy the recurrence relation
(omitting the second subindex $l=5, 6$)
\[
\fl 2\,\beta_1\,m\,b_m=\left[m(m-1)-16\,\mathcal{L}(\mathcal{L}+1)-15/4\right]\,b_{m-1}+16\,E\,b_{m-9}-16\,b_{m-17}.
\]
If we choose, as in the preceding examples,
\[
c_{0,1}=c_{0,2}=1,\quad a_{0,3}=a_{0,4}=1, \quad b_{0,5}=b_{0,6}=1,
\]
we have for the denominators in Eqs. (\ref{vuno}) and (\ref{vdos})
\begin{eqnarray}
\mathcal{W}[w_3,w_4] = - \,\mathcal{W}[w_4,w_3] = 8,  \label{qcuatro}  \\
\mathcal{W}[w_5,w_6] = - \,\mathcal{W}[w_6,w_5] = -8\,\lambda^{1/2}.  \label{qcinco}
\end{eqnarray}
The numerators in the same equations (for $j=1, 2$) are given by
\begin{eqnarray}
\fl\mathcal{W}[w_j,w_3] = \sum_{L=0}^7\,2^{n+\delta_L^{(j,3)}}\,\Gamma (n+1+\delta_L^{(j,3)})\, \gamma_{8n+L}^{(j,3)},  \label{qseis}\\
\fl\mathcal{W}[w_j,w_4] =(-1)^n\,\sum_{L=0}^7\,\cos(\pi\delta_L^{(j,4)})\, 2^{n+\delta_L^{(j,4)}}\,\Gamma (n+1+\delta_L^{(j,4)})\, \gamma_{8n+L}^{(j,4)}, \label{qsiete} \\
\fl\mathcal{W}[w_j,w_5] =(4\lambda^{1/2})^{-n+\nu_j+1}\,\Gamma (n-\nu_j)\,\gamma_{-n}^{(j,5)}, \label{qocho} \\
\fl\mathcal{W}[w_j,w_6] =(-1)^{n-1}\,\cos(\pi\nu_j)\,(4\lambda^{1/2})^{-n+\nu_j+1}\,\Gamma (n-\nu_j)\,\gamma_{-n}^{(j,6)}, \label{qnueve}
\end{eqnarray}
where we have abbreviated
\begin{eqnarray}
\fl\delta_L^{(j,k)}=(\nu_j+\mu_k+L)/8\,, \qquad k=3, 4,  \label{sesenta} \\
\fl\gamma_m^{(j,k)}=\sum_{s=0}^\infty a_{s,k}\left(\alpha_{8,k}\,{c}_{m+s-7,j}-(m\! +\! 2s\! +\! 1\! +\! \nu_j\! -\! \mu_k)\,
{c}_{m+s+1,j}\right),\qquad k=3, 4, \label{suno} \\
\fl\gamma_m^{(j,l)}=\sum_{s=0}^\infty b_{s,l}\left(-\ \beta_{1,l}\,c_{m-s+2,j}+(2s-m-\nu_j)\,c_{m-s+1,j}\right), \qquad l=5, 6.  \label{sdos}
\end{eqnarray}

The ground state energies obtained by using our procedure are reported in Table 9. We have taken for $\lambda$ several values already considered by other authors, whose results are shown also in Table 9 in order to facilitate comparison. As it can be seen, our result are considerably more accurate than those obtained by numerical integration of the Schr\"odinger equation \cite{agui2}, although we do not reach, with our double precision FORTRAN codes, the impressive accuracy of the results of Buend\'{\i}a {\it et al.} \cite{buen} obtained with the analytic continuation method. Our procedure, however,  can provide with results with as many correct digits as desired if a sufficiently precise arithmetic is used. And, very important, the wave function is obtained in a very convenient form (asymptotic expansions and Laurent series) for algebraic manipulations like normalization or computation of expected values.
\begin{table}
\caption{Ground-state energy of the critical spiked oscillator of potential (\ref{qtres}) for several values of the intensity $\lambda$. The column headed by $E$ shows our results. For comparison, we show, in the column headed by $E_{\scriptstyle{\rm lit}}$, values taken from the literature, namely from Refs. \cite{buen} (superscript a) and \cite{agui2} (superscript b), obtained by the analytic continuation method and by numerical integration, respectively. The last column shows the index $\nu_1$ of one of the Floquet solutions. For the index of the other one we have taken $\nu_2=1-\nu_1$.}
\begin{indented}
\item[]
\begin{tabular}{llll}
\br
 $\lambda$  & $\qquad E$ & $\qquad E_{\scriptstyle{\rm lit}}$ & $\qquad \nu_1$ \\
\mr
0.001  & 3.00401125101 & 3.004011251013044\,$^{\mbox{a}}$ & 0.5$\,+\,i\,$0.244567376746E$-$04 \\
0.005  & 3.01914010728 & 3.019140107276879\,$^{\mbox{a}}$ & 0.5$\,+\,i\,$0.612956032070E$-$03 \\
0.01   & 3.03672947263 & 3.036729\,$^{\mbox{b}}$ & 0.5$\,+\,i\,$0.245895012676E$-$02 \\
0.05   & 3.15242944140 & 3.152429\,$^{\mbox{b}}$ & 0.5$\,+\,i\,$0.625334357268E$-$01 \\
0.1    & 3.26687302611 & 3.266873026113018\,$^{\mbox{a}}$ & 0.5$\,+\,i\,$0.248671350579E$+$00 \\
0.5    & 3.84855317229 & 3.848553\,$^{\mbox{b}}$ & 0.5$\,+\,i\,$0.196189243685E$+$01 \\
1      & 4.31731168925 & 4.317311689247366\,$^{\mbox{a}}$ & 0.5$\,+\,i\,$0.288463702918E$+$01 \\
2      & 4.98613573609 &    & 0.5$\,+\,i\,$0.389836484474E$+$01 \\
5      & 6.29647263890 & 6.296472\,$^{\mbox{b}}$ & 0.5$\,+\,i\,$0.599616514206E$+$01 \\
10  & 7.73511110349 & 7.735111103489141\,$^{\mbox{a}}$ & 0.5$\,+\,i\,$0.813625698416E$+$01 \\
20  & 9.70940409621 &   & 0.5$\,+\,i\,$0.110399234356E$+$02 \\
\br
\end{tabular}
\end{indented}
\end{table}

\ack

The idea of applying our procedure to solve spiked oscillator problems was suggested to one of the authors (JS) by Prof. Aguilera-Navarro. Financial support from Comisi\'{o}n Inter\-mi\-nis\-te\-rial de Ciencia y Tecnolog\'{\i}a and of Diputaci\'on General de Arag\'on is acknowledged.

\appendix

 \section{Indices and coefficients of the Floquet solutions}

In Sec. 2 we have referred to the Floquet or multiplicative solutions (\ref{cinco}) of Eq. (3) and mentioned that their indices $\nu_j$ and coefficients $c_{n,j}$ must obey the infinite set of homogeneous equations (\ref{ocho}) and the condition (\ref{nueve}). (Along this Appendix we will omit, for brevity, the subindex $j$ in $\nu_j$ and $c_{n,j}$.)
As it has been already said, we face a nonlinear eigenvalue problem.
Algorithms to solve finite order problems of this kind have been discussed by Ruhe \cite{ruhe}.
Obviously, the condition (\ref{nueve}) requires
\begin{equation}
\lim_{n\to\pm\infty}\,|c_n|= 0\,.  \label{auno}
\end{equation}
This allows us to truncate our infinite problem (\ref{ocho}) by restricting the label $n$ to the interval $-\mathcal{M} \leq n \leq \mathcal{N}$,
 both $\mathcal{M}$ and $\mathcal{N}$ being positive integers large enough to guarantee
that the solution of the truncated problem does not deviate significantly from that of the original infinite one. Then, the Newton iteration method can be applied. It consists in moving from an approximate solution,
$\{\nu^{(i)}, c_n^{(i)}\}$, to another one,  $\{\nu^{(i+1)}, c_n^{(i+1)}\}$, by solving the system of equations
\begin{eqnarray}
\fl\Big(2n\! -\! 1\! +\! 2\nu^{(i)}\Big)c_n^{(i)}\Big(\nu^{(i+1)} - \nu^{(i)}\Big) +
\Big(n\! +\! \nu^{(i)}\Big)\Big(n\! -\! 1\! +\! \nu^{(i)}\Big)\,c_n^{(i+1)}
\nonumber \\ -\,\sum_{s=-2M}^{2N}g_s\,c_{n-s}^{(i+1)}  =  0, \qquad n=-\mathcal{M}, \ldots, -1, 0, 1, \ldots, \mathcal{N},
\label{ados} \\
\sum_{n=-\mathcal{M}}^\mathcal{N}{c_n^{(i)}}^*c_n^{(i+1)} =  1,  \label{atres}
\end{eqnarray}
that results, by linearization \cite{naund}, from (\ref{ocho}) and from the truncated normalization condition
\[
\sum_{n=-\mathcal{M}}^\mathcal{N}\left| c_{n}\right|^2 = 1.
\]
Obviously,
the values of $c_m^{(i)}$ with $m<-\mathcal{M}$ or $m>\mathcal{N}$ entering in some of Eqs. (\ref{ados}) should be taken
equal to zero, in accordance with the truncation done.  The iteration process is stopped when the
difference between consecutive solutions is satisfactory. The outcome may serve as starting point for a new iteration process with
larger values of $\mathcal{M}$ and $\mathcal{N}$, to check the stability of the solution.

The iteration process just described needs initial values $\{\nu^{(0)}, c_n^{(0)}\}$ not far from the true solution.
The two different values of $\nu$ can be obtained from the two eigenvalues
\begin{equation}
\lambda_j = \exp (2i\pi\nu_j)\,,    \qquad j=1, 2\,,   \label{acuatro}
\end{equation}
of the circuit matrix ${\bf C}$ \cite{waso} for the singular point at $z=0$.
The entries of that matrix can be computed by numerically integrating Eq. (\ref{tres}) on the unit circle, from $z=\exp (0)$ to $z=\exp (2i\pi)$, for two independent sets of initial values. If we consider two solutions, $w_a(z)$ and $w_b(z)$, obeying, for instance, the conditions
\begin{eqnarray}
w_a(\mbox{e}^0)=1,\qquad w_a^\prime(\mbox{e}^0)=0, \nonumber \\
w_b(\mbox{e}^0)=0,\qquad w_b^\prime(\mbox{e}^0)=1, \nonumber
\end{eqnarray}
then
\begin{eqnarray}
C_{11}=w_a(\mbox{e}^{2i\pi}), \qquad C_{12}=w_b(\mbox{e}^{2i\pi}), \nonumber \\
C_{21}=w_a^\prime(\mbox{e}^{2i\pi}), \qquad C_{22}=w_b^\prime(\mbox{e}^{2i\pi}), \nonumber
\end{eqnarray}
and
\begin{equation}
\nu=\frac{1}{2i\pi}\,\ln \left[\frac{1}{2}\left( C_{11}+C_{22}\pm\sqrt{\left(C_{11}\! -\! C_{22}\right)^2+4C_{12}C_{21}}\right)\right].
\label{acinco}
\end{equation}
The two signs in front of the square root produce two different values for $\nu$, unless the parameters
$g_s$ in Eq. (\ref{tres}) be such that $\left(C_{11}-C_{22}\right)^2+4C_{12}C_{21}=0$, in which case only one
multiplicative solution appears, any other independent solution containing logarithmic terms.
The ambiguity in the real part of $\nu$ due to the multivaluedness of the logarithm in the right
hand side of (\ref{acinco}) reflects the fact already mentioned that the indices $\nu$ are not uniquely defined.
Notice that
\[
\lambda_1\,\lambda_2=\det {\bf C}=\mathcal{W}[w_a,w_b]=1
\]
and, therefore,
\[
\nu_1+\nu_2=0\quad (\mbox{mod} \;1).
\]
This may serve as a test for the integration of Eq. (\ref{tres}) on the unit circle.

Although Eq. (\ref{acinco}) is exact, the $C_{mn}$ are obtained numerically and the resulting values of
$\nu$ may only be considered as starting values, $\nu^{(0)}$, for the Newton iteration process.
As starting coefficients $c_n^{(0)}$ one may use the solutions of the homogeneous system
\begin{equation}
\fl (n\! +\! \nu^{(0)})(n\! -\! 1\! +\! \nu^{(0)})\,c_{n}^{(0)} - \sum_{s=-2M}^{2N}g_s\,c_{n-s}^{(0)}=0\,, \qquad
  n= -\mathcal{M}, \ldots, -1, 0, 1, \ldots, \mathcal{N},   \label{aseis}
\end{equation}
with the already mentioned truncated normalization condition
\begin{equation}
\sum_{n=-\mathcal{M}}^\mathcal{N} |c_{n}^{(0)}|^2 = 1.  \label{asiete}
\end{equation}

\setcounter{section}{1}

\section{Wronskians of Floquet and Thom\'e solutions}

The connection factors of the Floquet solutions with the Thom\'e ones are given, by Eqs. (\ref{vuno}) and (\ref{vdos}), as quotients of two Wronskians. Those in the denominators  can be obtained immediately and were given in Eqs. (\ref{vtres}) and (\ref{vcuatro}). Direct computation of the Wronskians in the numerators must be discarded for the reasons pointed out at the end of Sec. 3. The purpose of this Appendix is to give a procedure to compute them. The idea is to find, for each one of the needed Wronskians,  two functions, one proportional to the other, the proportionality constant being that Wronskian. Comparison of analogous terms in the asymptotic expansions of those functions allows one to obtain the required Wronskian. This idea has already been exploited in the solution of the Schr\"odinger equation with a polynomial potential \cite{gom1,gom2}.

Let us consider the Wronskian of one of the Floquet solutions $w_j$ ($j=1, 2$), given in (\ref{cinco}), and one of the Thom\'e solutions at infinity $w_k$ ($k=3,4$), given in (\ref{seis}). We find convenient to introduce auxiliary functions
\begin{equation}
\fl u_{j,k}=\exp\left( -{\displaystyle{\frac{\alpha_{N,k}}{2N}}}\,z^{N}\right)w_{j},
\quad u_k=\exp\left( -{\displaystyle{\frac{\alpha_{N,k}}{2N}}}\,z^{N}\right)w_k, \qquad j=1, 2,  \quad  k=3, 4.  \label{buno}
\end{equation}
Obviously,
\begin{equation}
\mathcal{W}[u_{j,k},u_k]=\exp\left(-\frac{\alpha_{N,k}}{N}\,z^{N}\right) \mathcal{W}[w_{j},w_k]. \label{bdos}
\end{equation}
An asymptotic expansion of the left hand side of this equation can be calculated by using the definitions (\ref{buno}) and the expansions
(\ref{cinco}) and (\ref{seis}). It becomes
\begin{equation}
\fl\mathcal{W}[u_{j,k},u_k] \sim \left(\left(\alpha_{N,k}\,z^{N-1}+2\sum_{p=1}^{N-1}\alpha_{p,k}\,z^{p-1}\right)v_{j,k}-\frac{dv_{j,k}}{dz}\right)\mathcal{S}_k
+ v_{j,k}\,\frac{d\mathcal{S}_k}{dz},   \label{btres}
\end{equation}
where we have denoted
\begin{eqnarray}
v_{j,k}=\exp\left(\sum_{p=1}^{N-1}\frac{\alpha_{p,k}}{p}z^p\right)\,w_{j}\,, \qquad j=1, 2\,, \quad k=3, 4\,,  \label{bcuatro}  \\
\mathcal{S}_k = \sum_{m=0}^\infty a_{m,k}\,z^{-m+\mu_k}\,, \qquad k=3, 4\,. \label{bcinco}
\end{eqnarray}
For the newly introduced function $v_{j,k}$, a convergent Laurent expansion
\begin{equation}
v_{j,k} = \sum_{n=-\infty}^\infty \hat{c}_{n,j,k}\,z^{n+\nu_j}     \label{bseis}
\end{equation}
can be obtained as Floquet solution of the differential equation
\begin{eqnarray}
\fl -z^2\,\frac{d^2v_{j,k}}{dz^2}+2z\left(\sum_{p=1}^{N-1}\alpha_{p,k}\,z^p\right)\,\frac{dv_{j,k}}{dz}  \nonumber  \\
+ \left(\sum_{s=-2M}^{2N}g_s\,z^s+\sum_{p=2}^{N-1}(p-1)\alpha_{p,k}\,z^p-\left(\sum_{p=1}^{N-1}\alpha_{p,k}\,z^p\right)^2\right)v_{j,k} = 0\,.  \label{bsiete}
\end{eqnarray}
Then, by using Eqs. (\ref{bcinco}) and (\ref{bseis}) in (\ref{btres}), we obtain
\begin{equation}
\mathcal{W}[u_{j,k},u_k] \sim \sum_{n=-\infty}^\infty \gamma_n^{(j,k)}\,z^{n+\nu_j+\mu_k},   \label{bocho}
\end{equation}
where
\begin{eqnarray}
\fl \gamma_n^{(j,k)} =  \sum_{m=0}^\infty a_{m,k}\Bigg(\alpha_{N,k}\,\hat{c}_{n+m+1-N,j,k}
+ 2\sum_{p=1}^{N-1}\alpha_{p,k}\,\hat{c}_{n+m+1-p,j,k}
\nonumber  \\
\hspace{4cm} -\  (n+2m+1+\nu_j-\mu_k)\,\hat{c}_{n+m+1,j,k}\Bigg).    \label{bnueve}
\end{eqnarray}
The value of the $\mathcal{W}[w_{j},w_k]$ can be immediately obtained if we are able to write an asymptotic expansion of
$\exp\left(-\alpha_{N,k}\,z^{N}/N\right)$, in the right hand side of (\ref{bdos}), with the same powers of $z$ as the expansion
(\ref{bocho}) of the left hand side. With this purpose, we construct $N$ formal expansions
\begin{equation}
\mathcal{E}_{L}^{(j,k)}(z)=\sum_{n=-\infty}^{\infty}
\frac{\left(-\alpha_{N,k}\,z^{N}/N\right)^{n+\delta_{L}^{(j,k)}}}
{\Gamma(n+1+\delta_{L}^{(j,k)})}, \qquad L=0, 1, \ldots , N\! -\! 1,
\label{bdiez}
\end{equation}
of $\exp\left( -\alpha_{N,k}\,z^{N}/N\right)$. Such expansions are
but particular forms of the so called Heaviside's exponential
series
\begin{equation}
\exp(t)\sim\sum_{n=-\infty}^{\infty}\frac{t^{n+\delta}}{\Gamma(n+1+\delta)},
\label{bduno}
\end{equation}
introduced by Heaviside in the second volume of his {\em
Electromagnetic theory} (London, 1899) and which, as proved by
Barnes \cite{barn}, is an asymptotic expansion for arbitrary
$\delta$ and $|\arg (t)|<\pi$. Expansions of this kind have been
already used by Naundorf \cite{naun} in his treatment of the
connection problem, from which our method is a convenient
modification. It becomes evident that, for any set of constants
$\{\kappa_{L}^{(j,k)}\}$ ($L=0, 1, \ldots , N\! -\! 1$) satisfying the
restriction
\begin{equation}
\sum_{L=0}^{N-1} \kappa_{L}^{(j,k)} = \mathcal{W}[w_{j},w_k],  \label{bddos}
\end{equation}
one has from (\ref{bdos})
\begin{equation}
\mathcal{W}[u_{j,k},u_k]\sim\sum_{L=0}^{N-1} \kappa_{L}^{(j,k)}\,\mathcal{E}_{L}^{(j,k)}(z).
\label{bdtres}
\end{equation}
By choosing for the  $\delta_{L}^{(j,k)}$ in the expansions
$\mathcal{E}_{L}^{(j,k)}$ the values
\begin{equation}
\delta_{L}^{(j,k)}=(\nu_j+\mu_k+L)/N,  \label{bdcuatro}
\end{equation}
a comparison, term by term, can be done of the resulting expansion
in (\ref{bdtres}) with that in (\ref{bocho}). One obtains in this
way
\begin{equation}
\kappa_{L}^{(j,k)}\,\frac{\left(-\alpha_{N,k}/N\right)^{n+\delta_{L}^{(j,k)}}}
{\Gamma(n+1+\delta_{L}^{(j,k)})} = \gamma_{nN+L}^{(j,k)}\,, \label{bdcinco}
\end{equation}
for any positive integer $n$ large enough to satisfy
\[
|(n+\nu_j)(n-1+\nu_j)|>\sum_{s=-2M}^{2N}|g_s|\,.
\]
By substituting in (\ref{bddos}) the values of
$\kappa_{L}^{(j,k)}$ obtained from (\ref{bdcinco}) one has finally
\begin{equation}
\mathcal{W}[w_{j},w_k]=\sum_{L=0}^{N-1}\frac{\Gamma(n+1+\delta_{L}^{(j,k)})}
{\left(-\alpha_{N,k}/N\right)^{n+\delta_{L}^{(j,k)}}}\; \gamma_{nN+L}^{(j,k)},
\label{bdseis}
\end{equation}
where the minus sign in front of $\alpha_{N,k}$ is to be interpreted as an odd power of $e^{i\pi}$ or $e^{-i\pi}$ so as to have
$|\arg (-\alpha_{N,k}\,z^N)|<\pi$. In the physical problems, one is interested on the connection factors on the positive real semiaxis, that is, on the ray $\arg (z)=0$. In this case, $\arg (-\alpha_{N,3}\,z^N)=0$; then, Eq. (\ref{bdseis}) gives for one of the numerators in (\ref{vuno})
\begin{equation}
\mathcal{W}[w_{j},w_3]=\sum_{L=0}^{N-1}\frac{\Gamma(n+1+\delta_{L}^{(j,3)})}
{\left(|\alpha_{N,3}|/N\right)^{n+\delta_{L}^{(j,3)}}}\; \gamma_{nN+L}^{(j,3)}\,, \qquad j=1,2\,,
\label{bdsiete}
\end{equation}
and the connection factor $T_{j,4}$ can be obtained immediately from (\ref{bdsiete}) and (\ref{vtres}).
For the computation of the other connection factor, instead, one has to bear in mind the fact that the semiaxis $\arg (z)=0$ is a Stokes ray
for $T_{j,3}$. Actually, for $z$ on this ray, $|\arg (-\alpha_{N,4}\,z^N)|=\pi$ and the expansions (\ref{bdiez}) would not correspond to $\exp\left( -\alpha_{N,k}\,z^{N}/N\right)$. Following the common practice, we define $T_{j,3}$ on the Stokes ray by the average
\begin{equation}
T_{j,3}=\frac{1}{2}(T_{j,3}^{+} +T_{j,3}^{-})  \label{bdocho}
\end{equation}
of its values in the regions separated by that ray. This
corresponds to define
\begin{equation}
\mathcal{W}[w_{j},w_4] = \frac{1}{2}\left(
\mathcal{W}[w_{j},w_4]^{+} +
\mathcal{W}[w_{j},w_4]^{-}\right),
\label{bdnueve}
\end{equation}
an average of the Wronskians for $z$ slightly above and below the
positive real semiaxis. One obtains in this way
\begin{equation}
\fl\mathcal{W}[w_{j},w_4]=(-1)^n\,\sum_{L=0}^{N-1}\cos \left(\delta_L^{(j,4)}\,\pi\right)\,\frac{\Gamma(n+1+\delta_{L}^{(j,4)})}
{\left(\alpha_{N,4}/N\right)^{n+\delta_{L}^{(j,4)}}}\; \gamma_{nN+L}^{(j,4)}\,, \qquad j=1, 2\,,
\label{bveinte}
\end{equation}
and the expression of $T_{j,3}$ follows then from (\ref{bveinte}) and (\ref{vtres}).

The procedure to obtain the Wronskian of one of the Floquet solutions, $w_j$ ($j=1, 2$), and one of the Thom\'e solutions at the origin, $w_l$ ($l=5,6$) given in (\ref{siete}), is analogous to that just described. The auxiliary functions are now
\begin{equation}
\fl u_{j,l}=\exp\left( -{\displaystyle{\frac{\beta_{M,l}}{2M}}}\,z^{-M}\right)w_{j},
\quad u_l=\exp\left( -{\displaystyle{\frac{\beta_{M,l}}{2M}}}\,z^{-M}\right)w_l, \qquad j=1, 2,  \quad  l=5, 6.  \label{bvuno}
\end{equation}
Then,
\begin{equation}
\mathcal{W}[u_{j,l},u_l]=\exp\left(-\frac{\beta_{M,l}}{M}\,z^{-M}\right) \mathcal{W}[w_{j},w_l]. \label{bvdos}
\end{equation}
By using the definitions (\ref{bvuno}) and the expansions (\ref{cinco}) and (\ref{siete}) one obtains an asymptotic expansion of the left hand side, namely,
\begin{equation}
\fl\mathcal{W}[u_{j,l},u_l] \sim \left(\left(-\beta_{M,l}\,z^{-M-1}-2\sum_{q=1}^{M-1}\beta_{q,l}\,z^{-q-1}\right)v_{j,l}-\frac{dv_{j,l}}{dz}\right)\mathcal{S}_l
+ v_{j,l}\,\frac{d\mathcal{S}_l}{dz},   \label{bvtres}
\end{equation}
where we have denoted
\begin{eqnarray}
v_{j,l}=\exp\left(\sum_{q=1}^{M-1}\frac{\beta_{q,l}}{q}z^{-q}\right)\,w_{j}\,, \qquad j=1, 2\,, \quad l=5, 6\,,  \label{bvcuatro}  \\
\mathcal{S}_l = \sum_{m=0}^\infty b_{m,l}\,z^{m+\rho_l}\,, \qquad l=5, 6\,. \label{bvcinco}
\end{eqnarray}
A convergent Laurent expansion
\begin{equation}
v_{j,l} = \sum_{n=-\infty}^\infty \hat{c}_{n,j,l}\,z^{n+\nu_j}     \label{bvseis}
\end{equation}
for $v_{j,l}$ can be obtained as Floquet solution of the differential equation
\begin{eqnarray}
\fl -z^2\,\frac{d^2v_{j,l}}{dz^2}+2z\left(\sum_{q=1}^{M-1}\beta_{q,l}\,z^{-q}\right)\,\frac{dv_{j,l}}{dz}  \nonumber  \\
+ \left(\sum_{s=-2M}^{2N}g_s\,z^s+\sum_{q=1}^{M-1}(q+1)\beta_{q,l}\,z^{-q}-\left(\sum_{q=1}^{M-1}\beta_{q,l}\,z^{-q}\right)^2\right)v_{j,l} = 0\,.  \label{bvsiete}
\end{eqnarray}
From (\ref{bvtres}), by using Eqs. (\ref{bvcinco}) and (\ref{bvseis}), we obtain
\begin{equation}
\mathcal{W}[u_{j,l},u_l] \sim \sum_{n=-\infty}^\infty \gamma_n^{(j,l)}\,z^{n+\nu_j+\rho_l},   \label{bvocho}
\end{equation}
where
\begin{eqnarray}
\fl \gamma_n^{(j,l)} =  \sum_{m=0}^\infty b_{m,l}\Bigg(-\beta_{M,l}\,\hat{c}_{n-m+1+M,j,l}
- 2\sum_{q=1}^{M-1}\beta_{q,l}\,\hat{c}_{n-m+1+q,j,l}
\nonumber  \\
\hspace{4cm} +\  (-n+2m-1-\nu_j+\rho_l)\,\hat{c}_{n-m+1,j,l}\Bigg).    \label{bvnueve}
\end{eqnarray}
In order to obtain an asymptotic expansion of
$\exp\left(-\beta_{M,l}\,z^{-M}/M\right)$, in the right hand side of (\ref{bvdos}), with the same powers of $z$ as the expansion
(\ref{bvocho}) of the left hand side, we construct $M$ formal expansions analogous to those in (\ref{bdiez}),
\begin{equation}
\mathcal{E}_{L}^{(j,l)}(z)=\sum_{n=-\infty}^{\infty}
\frac{\left(-\beta_{M,l}\,z^{-M}/M\right)^{n+\delta_{L}^{(j,l)}}}
{\Gamma(n+1+\delta_{L}^{(j,l)})}, \qquad L=0, 1, \ldots , M\! -\! 1.
\label{btreinta}
\end{equation}
Let us now consider $M$ constants
$\{\kappa_{L}^{(j,l)}\}$ ($L=0, 1, \ldots , M\! -\! 1$) such that
\begin{equation}
\sum_{L=0}^{M-1} \kappa_{L}^{(j,l)} = \mathcal{W}[w_{j},w_l].  \label{btuno}
\end{equation}
Then,
\begin{equation}
\mathcal{W}[u_{j,l},u_l]\sim\sum_{L=0}^{M-1} \kappa_{L}^{(j,l)}\,\mathcal{E}_{L}^{(j,l)}(z).
\label{btdos}
\end{equation}
If we choose for the $\delta_{L}^{(j,l)}$ in (\ref{btreinta})  the values
\begin{equation}
\delta_{L}^{(j,l)}=(-\nu_j-\rho_l+L)/M\,,  \label{bttres}
\end{equation}
comparison of the resulting expansion in (\ref{btdos}) with that in (\ref{bvocho}) allows one to write
\begin{equation}
\kappa_{L}^{(j,l)}\,\frac{\left(-\beta_{M,l}/M\right)^{n+\delta_{L}^{(j,l)}}}
{\Gamma(n+1+\delta_{L}^{(j,l)})} = \gamma_{-nM-L}^{(j,l)}\,, \label{btcuatro}
\end{equation}
that, substituted in (\ref{btuno}), gives
\begin{equation}
\mathcal{W}[w_{j},w_l]=\sum_{L=0}^{M-1}\frac{\Gamma(n+1+\delta_{L}^{(j,l)})}
{\left(-\beta_{M,l}/M\right)^{n+\delta_{L}^{(j,l)}}}\; \gamma_{-nM-L}^{(j,l)},
\label{btcinco}
\end{equation}
where the minus sign in front of $\beta_{M,l}$ should be replaced by an odd power of $e^{i\pi}$ or $e^{-i\pi}$ such that
$|\arg (-\beta_{M,l}\,z^{-M})|<\pi$. For $z$ on the positive real semiaxis, one has
\begin{equation}
\mathcal{W}[w_{j},w_5]=\sum_{L=0}^{M-1}\frac{\Gamma(n+1+\delta_{L}^{(j,5)})}
{\left(|\beta_{M,5}|/M\right)^{n+\delta_{L}^{(j,5)}}}\; \gamma_{-nM-L}^{(j,5)}\,, \qquad j=1,2\,,
\label{btseis}
\end{equation}
and, analogously to Eq. (\ref{bveinte}),
\begin{equation}
\fl\mathcal{W}[w_{j},w_6]=(-1)^n\,\sum_{L=0}^{M-1}\cos \left(\delta_L^{(j,6)}\,\pi\right)\,\frac{\Gamma(n+1+\delta_{L}^{(j,6)})}
{\left(\beta_{M,6}/M\right)^{n+\delta_{L}^{(j,6)}}}\; \gamma_{-nM-L}^{(j,6)}\,, \qquad j=1, 2\,.
\label{btsiete}
\end{equation}
The connection factors $T_{j,5}$ and $T_{j,6}$ are then immediately obtained from Eqs. (\ref{vdos}) by using (\ref{vcuatro}), (\ref{btseis})
and (\ref{btsiete}).

 \setcounter{section}{1}

\end{document}